\documentclass[aps,prb,twocolumn,superscriptaddress,showpacs,floatfix]{revtex4}
\pdfoutput=1 
\usepackage{graphicx}
\usepackage{amsmath}
\usepackage{amssymb}
\usepackage{url}
\usepackage{xcolor}
\usepackage{color}

\begin{document}

\title{Behavior of test particles in the plasma sheath upon local bias and plasma switching}
\author{Gerald Schubert}
\affiliation{Institut f\"ur Physik, Ernst-Moritz-Arndt-Universit\"at
  Greifswald, Germany}
\affiliation{Regionales Rechenzentrum Erlangen, 
  Friedrich-Alexander-Universit\"at Erlangen-N\"urnberg, Germany}
\author{Ralf Basner}
\affiliation{Leibniz-Institute for Plasma Research and Technology (INP) Greifswald,
 Germany}
\author{Holger Kersten}
\affiliation{Institute for Experimental and Applied Physics, University of Kiel,
  Germany}
\author{Holger Fehske}
\affiliation{Institut f\"ur Physik, Ernst-Moritz-Arndt Universit\"at
  Greifswald, Germany}
\date{\today}


\begin{abstract}
  Equilibrating gravitation by electric forces, microparticles can be confined
  in the plasma sheath above suitably biased local electrodes.
  Their position depends on the detailed structure of the plasma sheath
  and on the charge that the particles acquire in the surrounding 
  plasma, that is by the electron and ion currents towards it.
  Bias switching experiments reveal how the charge and equilibrium
  position of the microparticle change upon altered sheath conditions.
  We observe similar particle behaviors also in the afterglow of the
  discharge for a persisting bias voltage on the electrode: damped
  oscillation into a new equilibrium or (accelerated) fall according
  to the bias.
\end{abstract}


\maketitle


\section{Introduction}

%
%
At the powered electrode of an asymmetric rf-discharge, self biasing
gives rise to a large potential drop, wide plasma sheath and strong
electric fields.
In contrast, the plasma sheath in front of grounded or additionally
biased electrodes is less pronounced and thinner.
The properties of such sheaths are, nevertheless, of fundamental
interest in view of plasma processing of
surfaces~\cite{BB02,KTFBDQWH05} and plasma chemistry.~\cite{OKS01}
Selective biasing of the substrate can be used to optimize the
electron and ion impact energies and directions.~\cite{LL05}

%
%
Monitoring plasma characteristics in the sheaths is challenging,
since common diagnostics, such as Langmuir-probes usually fail.
This is because the macroscopic tip of a Langmuir-probe significantly
alters the local plasma characteristics and cannot resolve density and
potential gradients in the narrow sheath with adequate precision.
In order to overcome these limitations, alternative diagnostics like
optical spectroscopy~\cite{CHLDR98,CLKD05} or microparticles as
probes~\cite{AGSRTM03,BSLSFK09} are used.
Those provide an efficient, nearly non-invasive tool 
for the local diagnostic of the plasma parameters.
%
In view of their practical use, a calibration is necessary to discern
their own charge dynamics from their capability to reflect the
properties of the surrounding plasma.

%
%
Upon immersion in a plasma, microparticles get charged by electrons
and ions that accumulate on their surface.
Since the electron mobility exceeds those of the ions, the particle
charge is in general negative, depending on the local
conditions of the surrounding plasma.
Albeit, it has been stressed that under certain conditions,
such as in a discharge afterglow, microparticles in the bulk plasma
can even acquire positive charges.~\cite{CMBS06}
While the plasma is operational, particle recharging takes place on
longer time scale than the almost instantaneous reaction of the plasma
to slightly modified external conditions.
Then, the particle charge constantly remains in equilibrium 
with the surrounding electrons and ions.
Drastic changes in the operating conditions, however, may perturb the
time scale ordering and hinder an equilibration of the particle
charge.

%
%
In this work, we investigate microparticles which are captured in the
plasma sheath and their reaction to changes in the operating
conditions of the plasma.
Thereby, we focus on switching off the plasma while the lower 
discharge electrode is grounded or additionally biased as well as 
changing the bias voltage during operation of the plasma.
The experimental results are discussed in view of a theoretical model
that combines basic equations with input data from particle-in-cell (PIC)
simulations.

\section{Experimental setup}

The PULVA-INP setup~\cite{TBWK08} consists of an asymmetric, capacitively coupled 
rf-plasma in argon, working at neutral gas pressures $p_{\text{Ar}}$ from 0.1 Pa 
to 100 Pa.
The rf-power ($P_{\text{rf}}[\text{W}]=5-100$) is supplied by the upper,
powered electrode at a frequency of $\nu_{\text{rf}}=13.56\,\text{MHz}$
and amplitudes $\Phi_{\text{rf}}$ up to 1000 V.
In dependence on the external parameters $P_{\text{rf}}$ and $p_{\text{Ar}}$,
the obtained characteristics for the pristine argon plasma are
electron densities $n_{\text{e}}[ \text{cm}^{-3}]=10^9-10^{11}$, electron 
energies $k_{\text{B}}T_{\text{e}}[\text{eV}]=0.8-2.8$ and plasma potentials with 
respect to the ground of $\Phi_{\text{p}}[\text{V}]=20-30$.~\cite{TTBHGK06}
The overall plasma characteristics are monitored by Langmuir probe 
and  plasma monitor measurements.

\begin{figure}
  \centering
  \includegraphics[width=0.8\linewidth,clip]{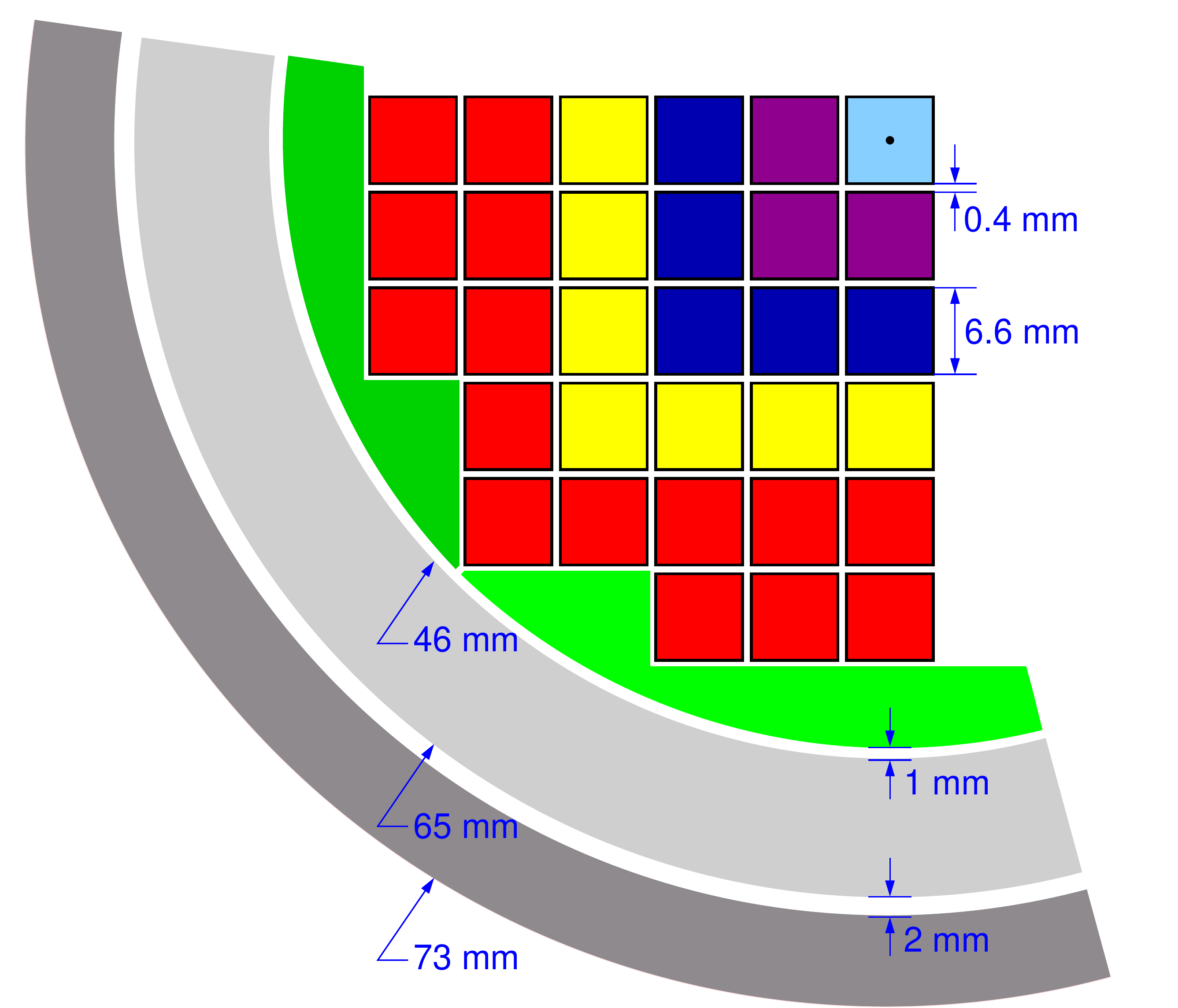}
  \caption{(Color online)
    Setup of the adaptive electrode (only one quarter is shown).
    From the center outward: square pixels, fitting segments (green), 
    ring (light grey) and ground shield (dark grey). The spacings 
    between all constituents are isolating.}
  \label{fig:AE}
\end{figure}

The key feature of the experimental setup is the lower,
so called `adaptive' electrode (AE).
It consists of 101 square electrode segments (pixels) with a linear 
extension of $6.6\,\text{mm}$ each, separated by thin
($0.4\,\text{mm}$) isolating gaps (see Fig.~\ref{fig:AE}).
In addition, four larger segments fit the pixel geometry to the surrounding 
ring and ground shield.
All 105 electrode pixels can be biased individually or in groups
by an external dc-voltage of up to $\Phi_{\text{bias}}=\pm100\,\text{V}$.
%
%
%
The selective application of bias voltages to some pixels allows for 
studying spatial and temporal changes of the plasma sheath.\cite{AGSRTM03,TBWK08}
The small extension of the pixels as compared to the remaining grounded 
electrode guarantees that the applied bias only locally influences the
plasma sheath but leaves the overall discharge conditions unaltered.

All experiments in this work refer to a plasma operating power 
$P_{\text{rf}} = 10\,{\text W}$ and neutral gas pressure 
$p_{\text{Ar}}=5\,\text{Pa}$ unless stated differently.
The microparticles used as probes are melamine-formaldehyde (MF)
spheres with diameter $d = 9.6\,\mu\text{m}$.
With a mass density of
$\rho_{\text{m}}=1.51\times10^{3}\,\text{kg}\,\text{m}^{-3}$ for MF
this results in a particle mass of $m=7\times10^{-13}\,\text{kg}$.
Illuminated by a laser fan (532 nm), their positions are recorded by a
fast CCD camera (up to 2000 frames per second).

\section{Results and Discussion}

In this work we report on three different experiments.
Each one describes a particular way in which the equilibrium
conditions for a levitating particle in the plasma sheath are
perturbed.
The initial position of the particle is determined by an
equilibrium between gravitational and electrostatic forces.
Neutral drag and thermophoresis effects are negligible. 
The effect of ion drag forces will be discussed in the experiments for
which it is relevant.

\subsection{Plasma afterglow -- unbiased AE}

\begin{figure}[t]
\centering\includegraphics[width=\linewidth,clip]{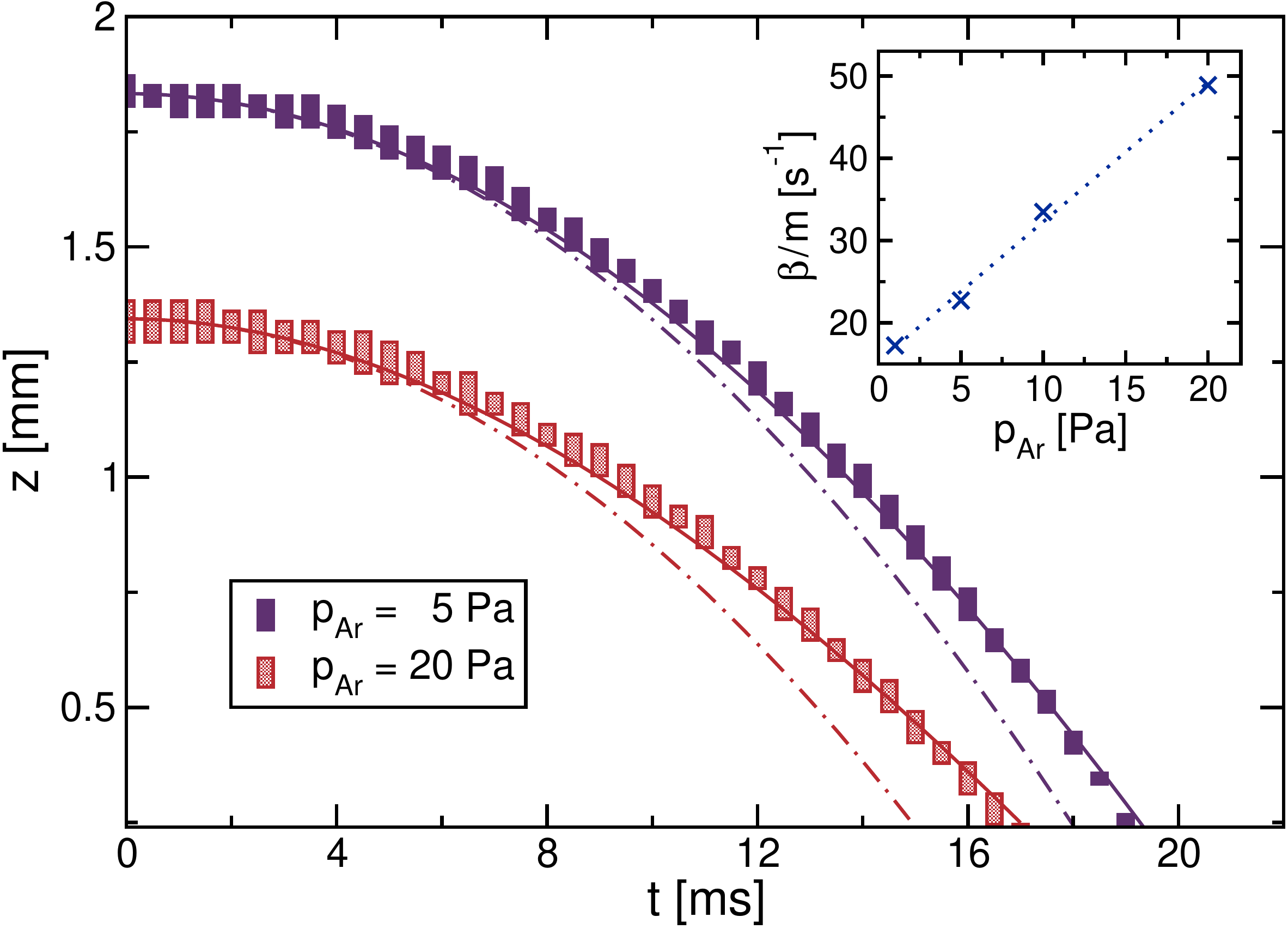}
\caption{(Color online) Main panel: Measured vertical particle 
  position after switching off the plasma for different neutral gas 
  pressure $p_{\text{Ar}}$ (boxes). 
  Theoretical trajectories for freely falling [retarded by friction
  according to  (\ref{eq:traj_retarded})] particles are given by 
  dashed-dotted [solid] lines.
  Inset:
  Crosses give the friction coefficient $\beta/m$ as extracted from 
  fitting (\ref{eq:traj_retarded}) to the experimental data.
  The dotted line is a guide to the eye.
\label{fig:fall_off_nobias}
}
\end{figure}

In this experiment we start from an initial configuration where the
MF-particle is trapped above the unbiased center pixel by a confining
potential of $-5\,\text{V}$ on the surrounding pixels of the AE.
At $p_{\text{Ar}}=5\,\text{Pa}$, the sheath width is about
$3\,\text{mm}$ and the equilibrium position $z_0=1.8\,\text{mm}$.
Switching off the plasma, the electrostatic force no longer compensates
gravity and the particle drops onto the AE.
As compared to a freely falling particle, the drop is markedly retarded 
(see Fig.~\ref{fig:fall_off_nobias}).

\label{sec:off_unbiased}

Along with the residual electric field, which decays over some
$\mu\text{s}$, also the directed ion current towards the AE vanishes
and ion drag can be neglected.
%
The only forces relevant on the time scale of the fall ($\text{ms}$)
are gravitation $\vec{F}_{\text{g}} = -mg\vec{e}_z$ and neutral drag force
$\vec{F}_{\text{n}} = -\beta\vec{\dot{z}}= -\beta \dot{z}\vec{e}_z$.
Here the damping constant $\beta$ depends on the local density of the
background gas, which we assume to be constant in time.
Then the equation of motion for the particle reads
$\ddot{z} + \frac{\beta}{m}\dot{z} + g = 0$.
The solution of this equation, subject to the initial conditions $\dot{z}(t=0) = 0$
and $z(t=0)=z_0$ is given by
\begin{equation}
  z(\tau) = z_0 + \frac{m^2 g}{\beta^2}\left(1-e^{-\tau}-\tau\right)\,, 
  \label{eq:traj_retarded}
\end{equation}
where we introduced the normalized time $\tau=\beta t/m$.
Depending on the argon gas pressure, the values of $\beta/m$ obtained
by a least squares fit of (\ref{eq:traj_retarded}) to the experimental
data are in the range of $20\text{s}^{-1}$ to$50\text{s}^{-1}$ .
This so called the Epstein friction
coefficient~\cite{Ep24} can be related to the neutral gas pressure by
\begin{equation}
  \frac{\beta}{m} = \delta\frac{8}{\pi}
  \frac{p_{\text{Ar}}}{\rho_{\text{m}} v_{\text{Ar}}^{\text{th}} d/2}\,,
  \label{eq:friction}
\end{equation}
where $v_{\text{Ar}}^{\text{th}}$ is the thermal velocity of the argon gas atoms
and $\delta$ is a parameter in the order of $1-1.5$ accounting for how the 
gas atoms are deflected from the particle surface.~\cite{NI08}
Assuming a diffuse reflection,
we obtain a theoretical value of $\beta/m\sim30\,\text{s}^{-1}$ for
$p_{\text{Ar}}=20\,\text{Pa}$ at room temperature, which
is in good accordance with the values determined by the measurement.
Furthermore, to describe our experimental data it is not necessary to
adapt $\beta$ to the decaying neutral gas density in the discharge
afterglow, as discussed in Ref.~\onlinecite{CMBS06}.
Within our accuracy a constant value of $\beta$ is sufficient.

Due to the vanishing electric field this experiment does not provide
any information on the value or variation of the particle charge.
In order to obtain information on this aspect, we have to resort to a
slightly modified setup, retaining a static dc-bias at some pixels of
the AE.

\subsection{Plasma afterglow -- biased AE}

\begin{figure}
\centering\includegraphics[width=\linewidth,clip]{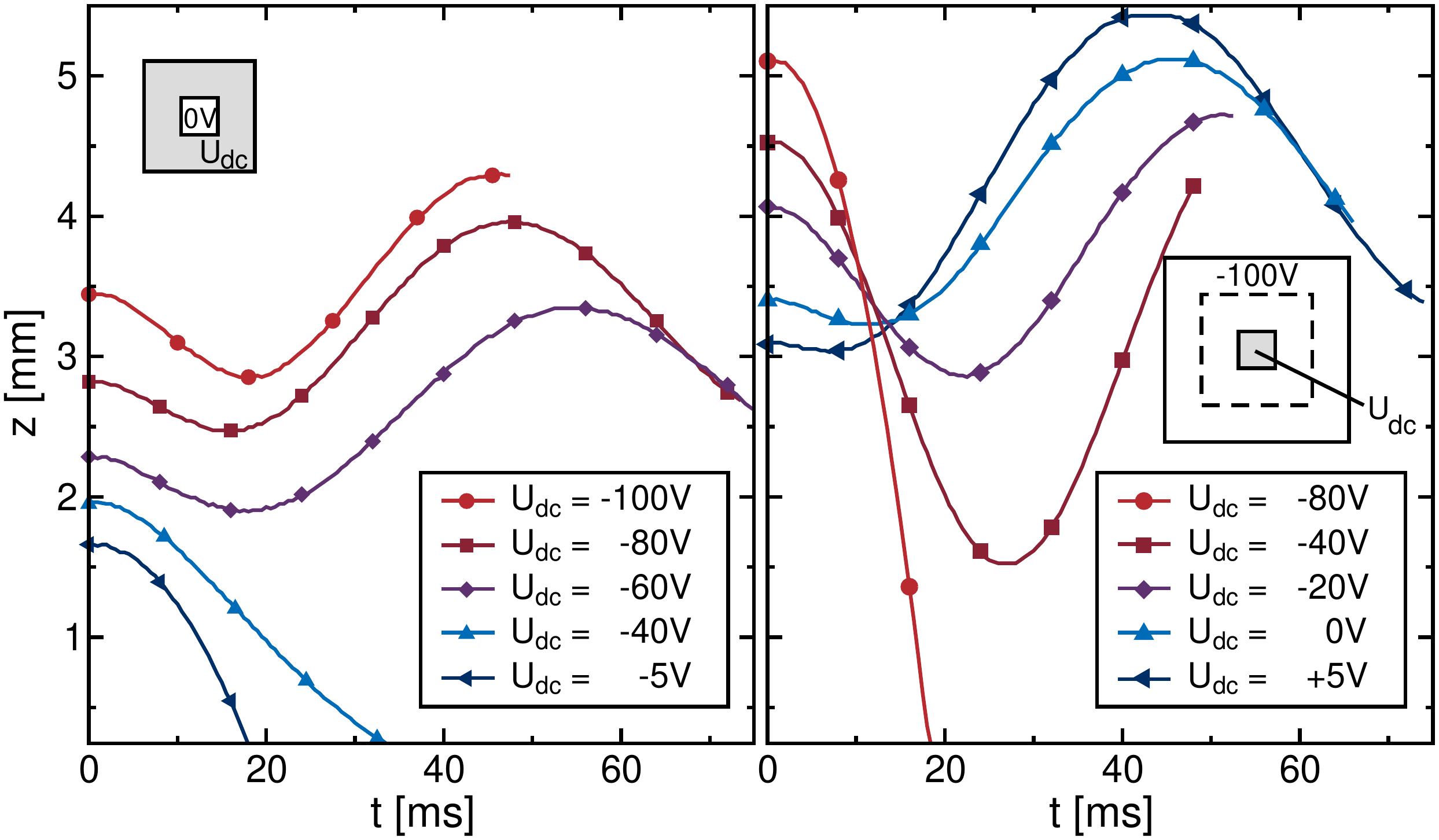}
\caption{(Color online) Measured vertical particle position after
  switching off the plasma while retaining dc-biases on selected
  segments of the AE. Left panel: Bias voltages 
  $U_{\text{AE}} = (0, U_{\text{dc}}, 0)$, where
  the three numbers refer to the pixel biasing from the center of the
  AE outward -- (light blue, violet, dark blue) with pixel colors referring to
  Fig.~\ref{fig:AE}. The rest of the AE is grounded.
  Right panel:  Bias voltages $U_{\text{AE}} = 
  (U_{\text{dc}}, -100\,\text{V}, -100\,\text{V})$. Time resolution of the
  measurement is $0.5\,\text{ms}$ and symbols are given to distinguish
  the curves.  Error bars for the positions are comparable to the
  symbol size.
}
\label{fig:fall_off_dcbias}
\end{figure}
Extending the previous experiment, we start from various bias configurations
of the AE by which we initially trap a particle.
Clearly, the biased pixels locally warp the plasma sheath.
Thus, the equilibrium positions (and presumably also the charges) of
the particles differ for each case.
The configurations for which the results are given in
Fig.~\ref{fig:fall_off_dcbias} differ in the number of biased pixels
-- $(3\times3)$ for the left and $(5\times5)$ for the right panel,
respectively.
The applied bias voltages exhaust the available range provided by the AE, 
resulting in weak to strong confinement.
For weak confinement ($|U_\text{dc}| < 60\,\text{V}$) we recover
in the left panel of Fig.~\ref{fig:fall_off_dcbias} the falling behavior
observed in the previous experiment.
Only for stronger negative bias ($|U_\text{dc}| \ge 60\,\text{V}$) the
persisting electric field is strong enough to keep the particle
hovering above the AE.
Note that the lengths of the recorded time intervals are limited by the
particle horizontally leaving the camera focus.
Their levitating state persists over several seconds.
The particle trajectories in the right panel of
Fig.~\ref{fig:fall_off_dcbias} are shifted to markedly larger
distances from the AE.
This reflects the more pronounced widening of the plasma sheath by
the larger area of strongly biased pixels.
As long as the surrounding pixels are on a sufficiently negative bias,
the particle levitation even persists for a moderate positive voltage
on the center pixel.
For a large negative bias ($U_\text{dc} = -80\,\text{V}$), however,
switching off the plasma causes an abrupt drop of the particle, which
is even accelerated as compared to a free fall.

\begin{figure}
\centering\includegraphics[width=0.36\linewidth,clip]{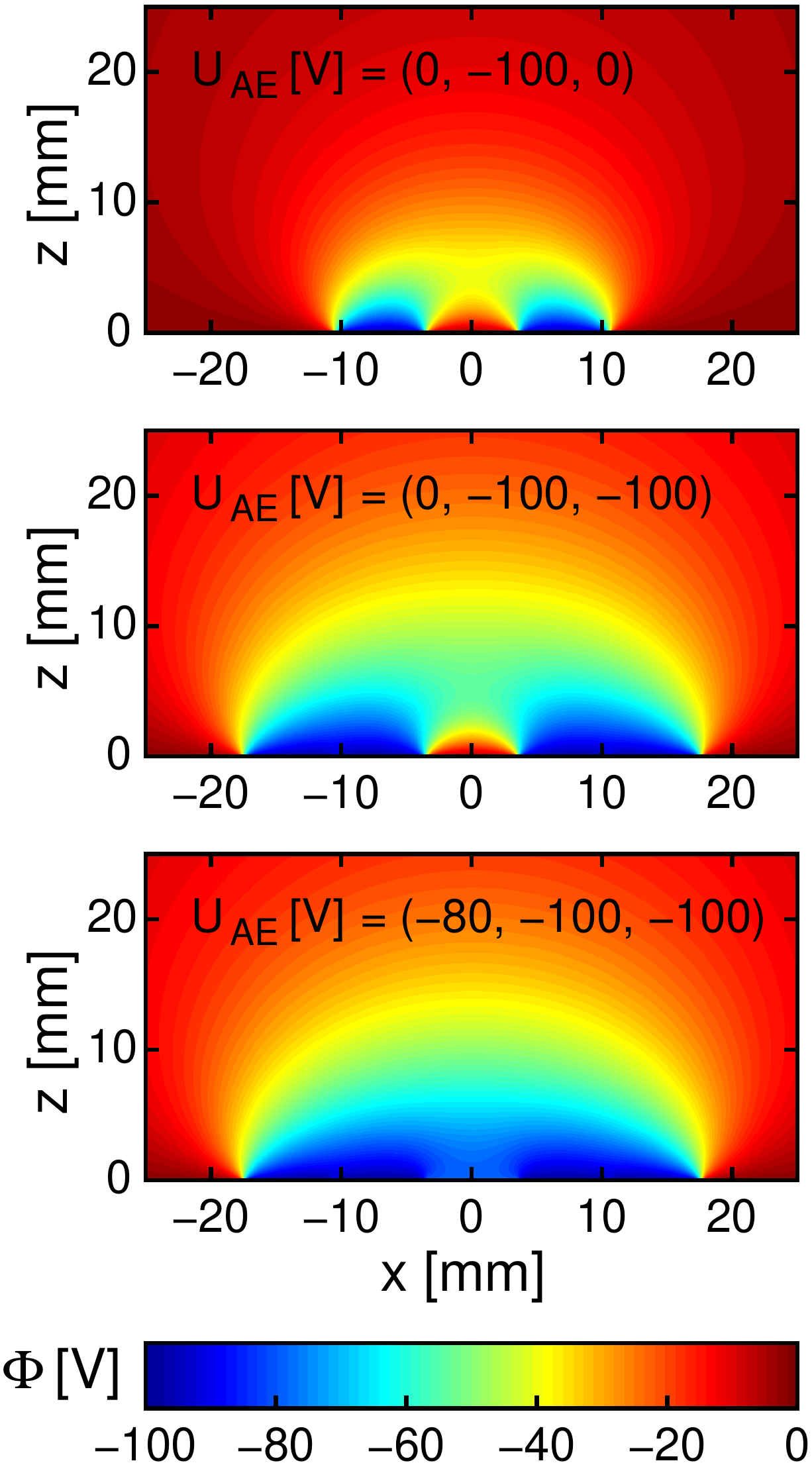}
\hfill
\centering\includegraphics[width=0.61\linewidth,clip]{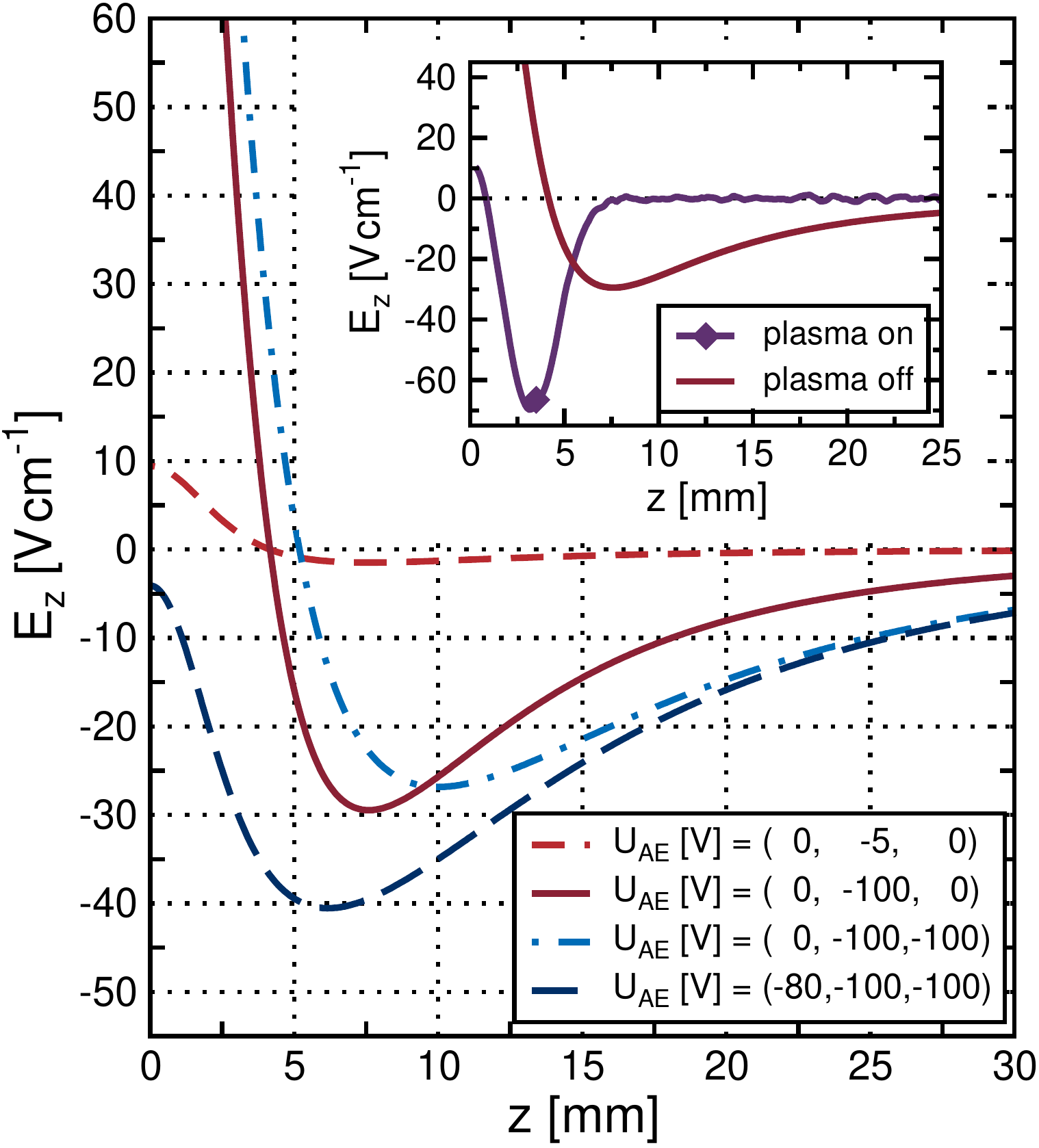}
\caption{(Color online) Left panel: Vertical cut through the
  electrostatic potential distribution above the AE in absence of any
  plasma. The shown region depicts
  only a fraction of the calculation grid ($n_x\times
  n_z=1400\times2000$ grid points corresponding to $7\,\text{cm}\times
  10\,\text{cm}$, $x\ge0$).  Right panel: Vertical component of the
  electric field on the symmetry axis as extracted from the potential
  data. In the inset the electric field for
  $U_{\text{AE}}[V]=(0,-100,0)$ before and after switching off the
  plasma is given. The data for operating plasma result from a PIC
  simulation. The diamond indicates the measured initial equilibrium
  position of the particle.}
\label{fig:Vertical_component}
\end{figure}
In order to explain the observed behavior, a first idea would
be to assume that on the observed time scales the plasma has
completely decayed, i.e. all electrons and ions have already recombined.
Then, the particle trajectory corresponds to the motion of a
charged sphere between two parallel plates, that are the powered
electrode and the AE.
Using the superposition principle, we may calculate the electrostatic force
acting on the particle in two steps.
First, the self-force for a charged particle between two parallel,
grounded planes is calculated analytically by means of an infinite
series of image charges~\cite{Ja98c} (see Appendix A).
This contribution is, however, negligible for distances $z \gtrsim 0.1\,\text{mm}$.
Second, we numerically solve the boundary value problem and calculate the 
potential due to the biased electrode pixels.
To this end, we reduce the simulation volume to a two-dimensional,
cylindrically symmetric domain by smearing out the individual pixels
to rings, and use routines from FISHPACK.~\cite{FISHPACK}
In vertical direction the simulation volume covers the whole range between
both electrodes.
Radially, the simulation domain is such large that the
boundaries do not affect the results near the center of the discharge.
At $r_{\text{max}}$ we used vanishing radial field boundary conditions.

Due to rotational symmetry, the biasing of outer lying pixels
influences the electrostatic potential above the center pixel to the
point of generating a region with negative potential at finite $z$
(left panel of Fig.~\ref{fig:Vertical_component}).
Above an unbiased center pixel this gives rise to a potential structure
with a minimum at $z_0$.
For strong negative biasing of the center pixel,
$U_{\text{AE}}[V]=(-80,-100,-100)$, the potential along the symmetry axis 
monotonically increases with $z$.
The sign change of the electric field at $z_0$ implies that a negatively
charged particle may only remain hovering for $z>z_0$ since there
gravity and electrostatic force are anti\-parallel (right panel of
Fig.~\ref{fig:Vertical_component}).
As compared to the case of operating plasma, the electric field after
the switching off is markedly reduced (inset of the right panel
of Fig.~\ref{fig:Vertical_component}).
Therefore, keeping up the force equilibrium requires an increasing 
of the particle charge.
Whether the particle remains hovering or drops down onto the AE
depends on its ability to acquire enough negative charges on a
sufficiently short time scale.
For systems with markedly higher operating pressures the relevant
stages of a plasma decay have been identified by Cou{\"e}del et
al.~\cite{CMBS06}, focussing on the bulk properties.
The properties of the sheath in the afterglow of a pulsed inductively
coupled rf-plasma in hydrogen have been addressed by Osiac et
al.~\cite{OSOHPTGC07}
For the locally biased plasma sheath the following
aspect comes into play.
After the switching off the periodic flooding of the sheath with
electrons during each rf-period ceases, but electrons and ions
continuously diffuse from the bulk into the sheath.
Here, the flow of both species is controlled by the applied local bias
voltages.
The charge equilibrium in the sheath will vary only slightly
as long as enough electrons and ions from the bulk enter the sheath
to compensate losses at the wall.
A particle adapts its charge to the altered electron- and
ion density in surrounding on time scales of some
$1-10\,\mu\text{s}$.\cite{MS06}
Since this time scale is faster than the changes of the plasma sheath
in the discharge afterglow, the particle charge constantly remains
in equilibrium with the charge densities in its surrounding.
In order to explain the accelerated fall for $U_{\text{AE}}[V]=(-80,-100,-100)$
consider that due to the strong negative bias no electrons at all will
be able to penetrate into the sheath above the center pixel anymore.
Therefore, the exclusive presence of ions in the surroundings of the
particle will drastically reduce its negative charge.
One might even have to consider a positive recharging since gravity
and ion drag force~\cite{BKFOC92,VT98,VC00} are too weak to account
for the observed behavior alone.
The individual contributions to the total force causing the
accelerated fall will be analyzed for the bias switching experiment in
the next section.

\subsection{Persisting plasma -- bias switching}
\begin{figure}
\centering\includegraphics[width=\linewidth,clip]{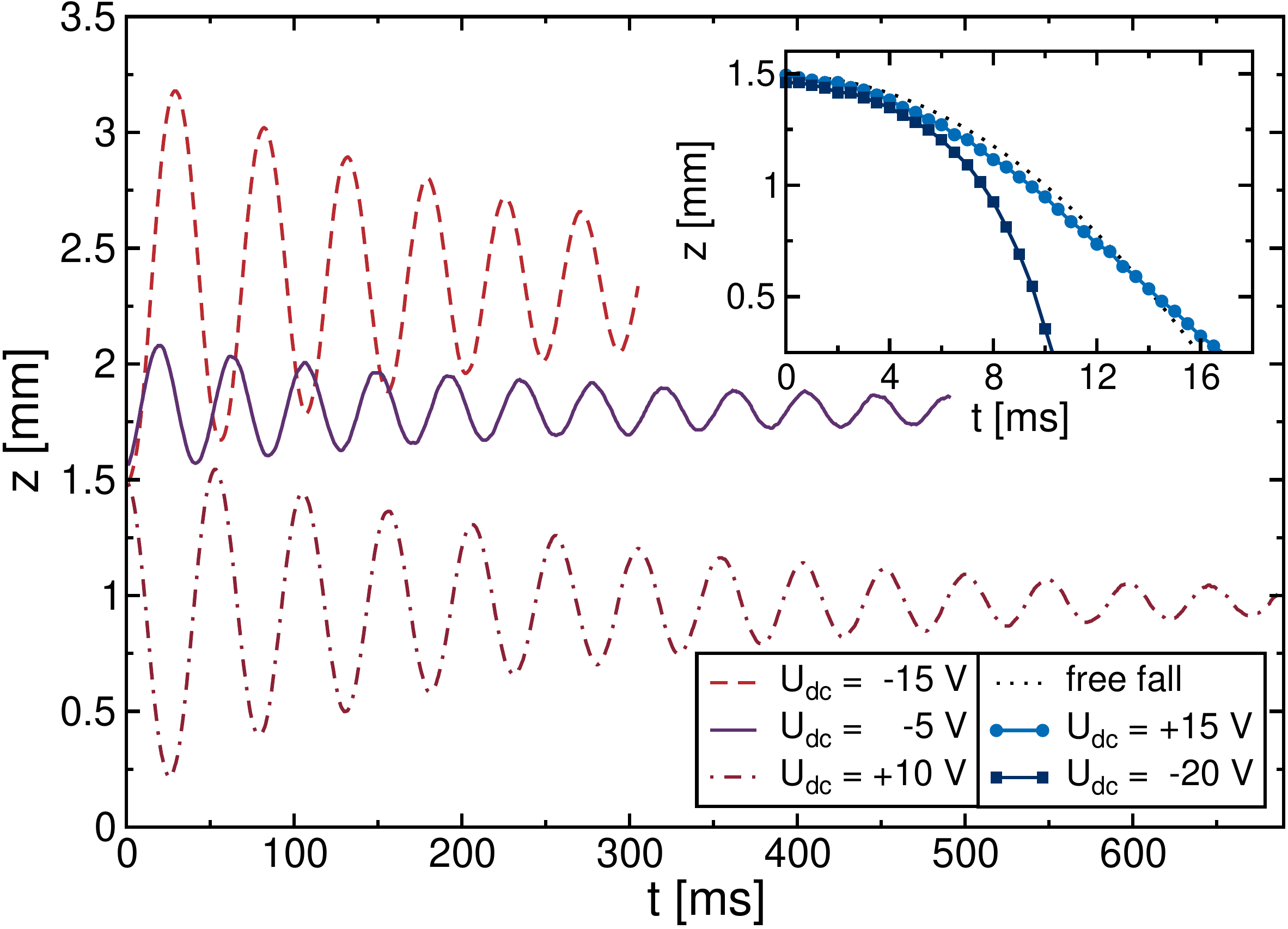}
\caption{(Color online) Measured vertical particle position upon bias
  switching while the plasma is kept operational.  The initial
  confining potential is $U_{\text{AE}}[\text{V}]=(0,-5,-50)$, which
  is switched to the final configuration
  $U_{\text{AE}}=(U_{\text{dc}},U_{\text{dc}},-50\,\text{V})$. In the
  inset the time scale is magnified to resolve the dynamics of those
  particles that drop onto the AE. For comparison the trajectory of a
  freely falling particle is given.  }
\label{fig:switch}
\end{figure}

While switching off the plasma certainly alters the equilibrium 
conditions for the particle drastically,
the changes in densities and fields occur on a moderate time scale.
This is in contrast to the situation where the AE-pixel biasing is
switched during operation of the plasma. 
Here, the local plasma sheath adapts within $1-5$ rf-cycles
to the new biasing configuration and thus has an instantaneous 
impact on the charge and force equilibrium of the particle. 
%
%
In principle, a particle may be trapped into a stable equilibrium
position for all indicated biasing configurations in
Fig.~\ref{fig:switch}, provided that the relaxation is performed
smoothly.
Upon an abrupt change from an initial configuration, however, the
particle drops onto the AE outside a rather narrow region of bias
variations (see inset of Fig.~\ref{fig:switch}).
This holds for either positive or negative voltages.
Interestingly, again a strong negative bias leads to an acceleration
of the particle towards the AE which is not explicable solely by
gravity.
For moderate changes in the biasing configuration the particle reaches
its new equilibrium position through a damped oscillatory motion.

\begin{figure}
  \centering
  \includegraphics[width=\linewidth,clip]{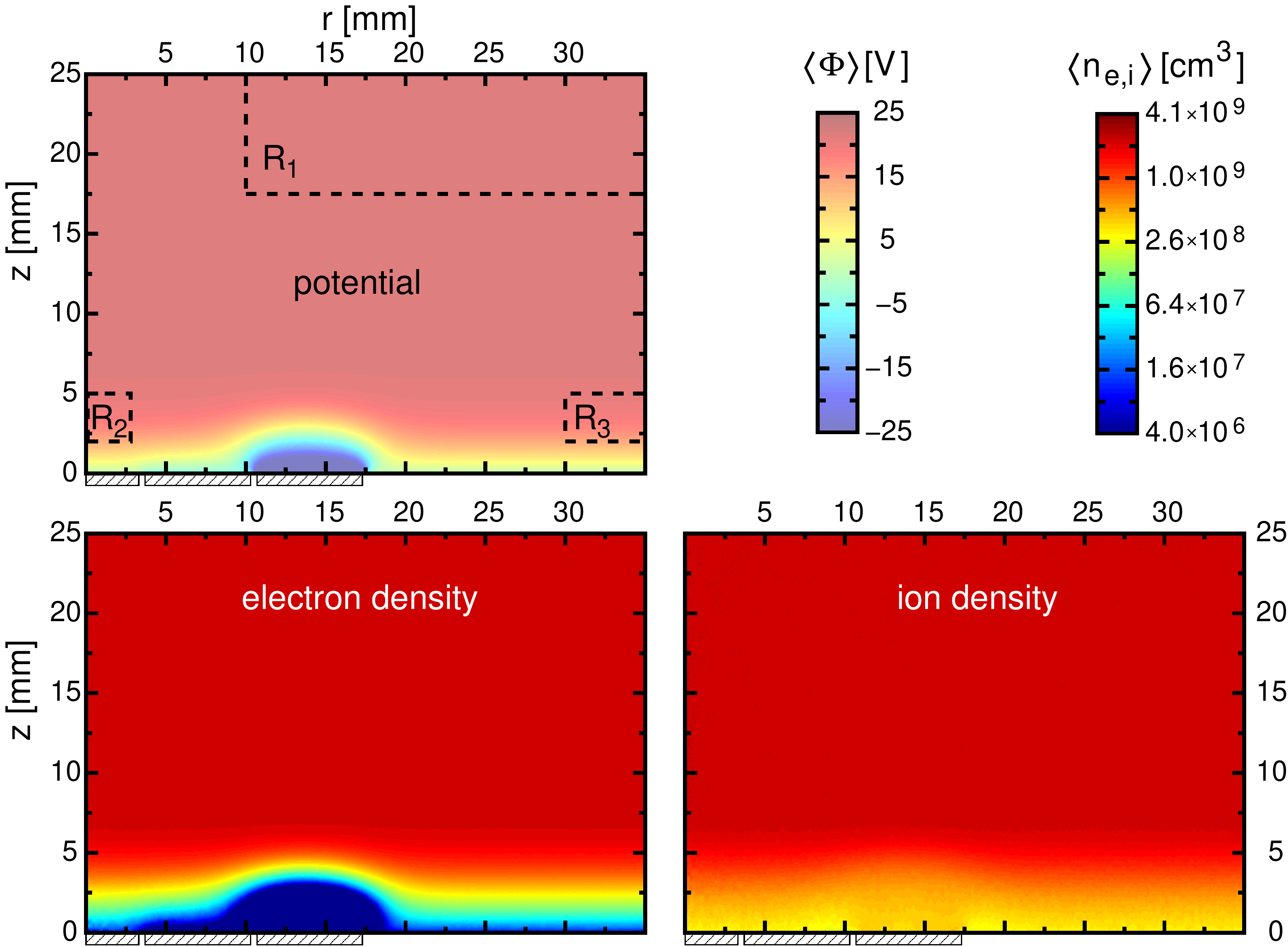}
  \caption{(Color online) PIC-results for the time-averaged potential
    as well as electron and ion density. Only part of the total
    simulation volume is shown. The AE-pixels are indicated
    by hatched rectangles and the biasing corresponds to the initial
    configuration in Fig.~\ref{fig:switch}.  In the upper panel $R_1 -
    R_3$ indicate the diagnostic regions for which the local electron
    energy probability distributions are given in
    Fig.~\ref{fig:local_EEPF}.}
  \label{fig:Diagnostic_regions}
\end{figure}

\begin{figure}
\centering\includegraphics[width=\linewidth,clip]{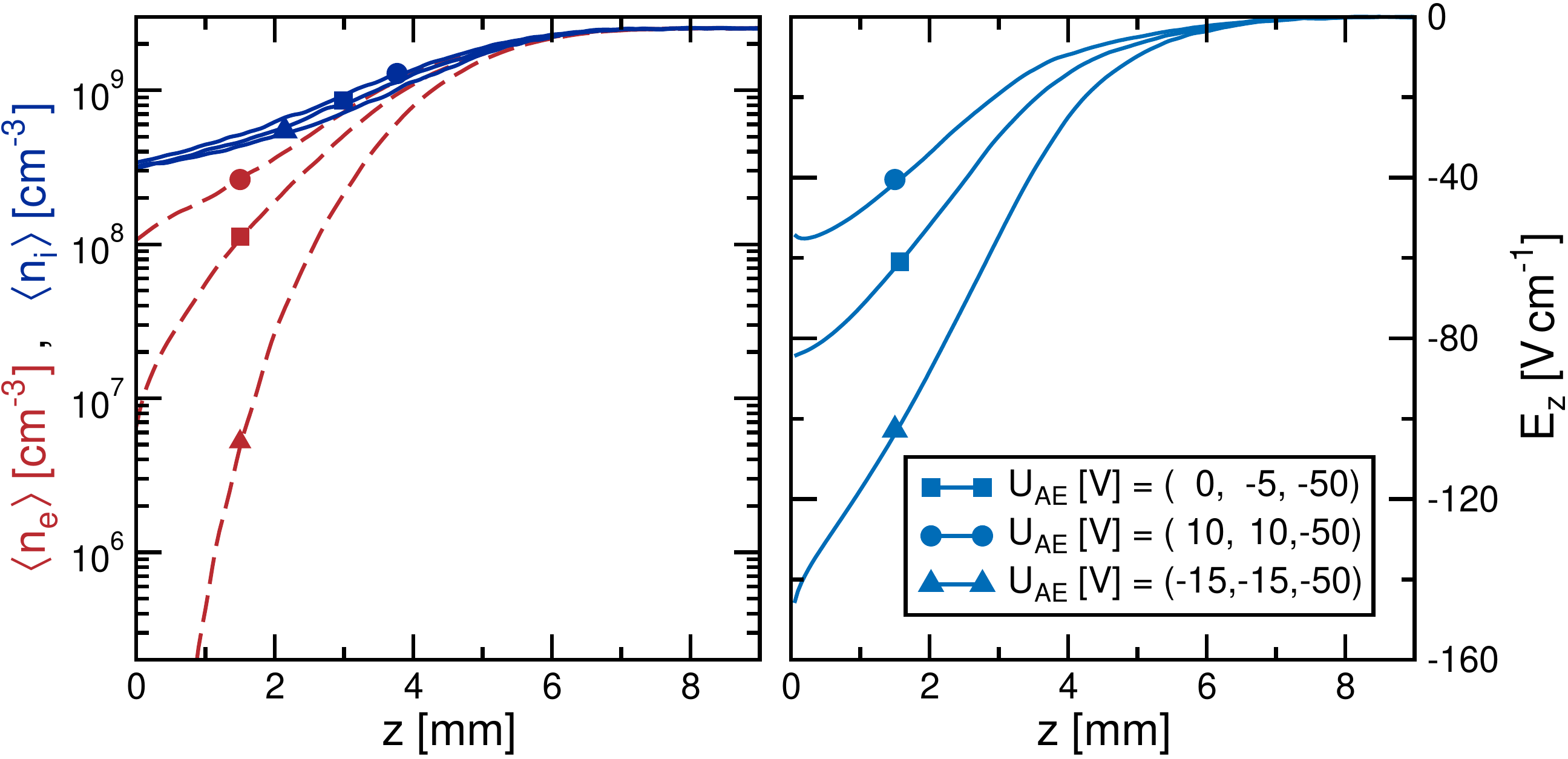}
\caption{Influence of the pixel biasing on electric field as well as
  electron and ion density on the symmetry axis of the discharge.
  Results obtained by PIC simulation.}
 \label{fig:change_sheath}
\end{figure}

In order to describe the different behaviors quantitatively, we
performed a PIC-simulation of the plasma discharge.
Special emphasis is put on the local modifications of the plasma 
sheath by the AE-pixels.
Details on the simulation can be found in Appendix~\ref{App_PIC}.

The initial AE-configuration in Fig.~\ref{fig:switch} only marginally
disturbs the plasma sheath above the center pixel.
Its primary purpose is to keep the particle in the camera focus.
As can be seen from the simulation results in
Fig.~\ref{fig:Diagnostic_regions}, both the electron and ion densities
near the symmetry axis agree with those in the unperturbed sheath
far away from the biased pixels.
In this respect, the results for $U_{\text{AE}}[\text{V}]=(0,-5,-50)$ in
Fig.~\ref{fig:change_sheath} are representative for an
unperturbed sheath.
While the two-dimensional color plots in Fig.~\ref{fig:Diagnostic_regions} 
provide a general overview of the sheath structure above 
the AE-pixels, Fig.~\ref{fig:change_sheath} allows for a quantitative 
analysis on the symmetry axis.
The time-averaged electron density is markedly reduced as compared to 
the ion density. 
Electrons may overcome the potential barrier of the plasma potential 
and enter the sheath only during a short fraction of an rf-period.
This occurs, when the potential on the powered electrode is sufficiently
negative to supply enough kinetic energy to the electrons.
Upon additional negative biasing of an AE-pixel, the necessary energy for
crossing the sheath increases.
This results in a further reduction of the electron density since
for most electrons the sheath becomes impenetrable.
Having lost their kinetic energy already after some part of the sheath, 
they remain trapped in the plasma.
Hence, above the negatively biased pixel,  parts of the sheath may be 
completely deprived of electrons.
Because of the continuity equation and acceleration of the ions towards
the AE, their (time-averaged) density also decreases in the sheath.
Their larger mass makes them less susceptible to further acceleration
by additional biasing and therefore all ion densities nearly coincide.
In contrast to switching off the plasma, which leads to wider final
potential structures and weaker fields, a negative bias may
drastically enhance the electric field in the sheath.
This enhancement results from the dominance of the increased potential 
difference over the effect of sheath widening.

\begin{figure}
  \centering
  \includegraphics[width=\linewidth,clip]{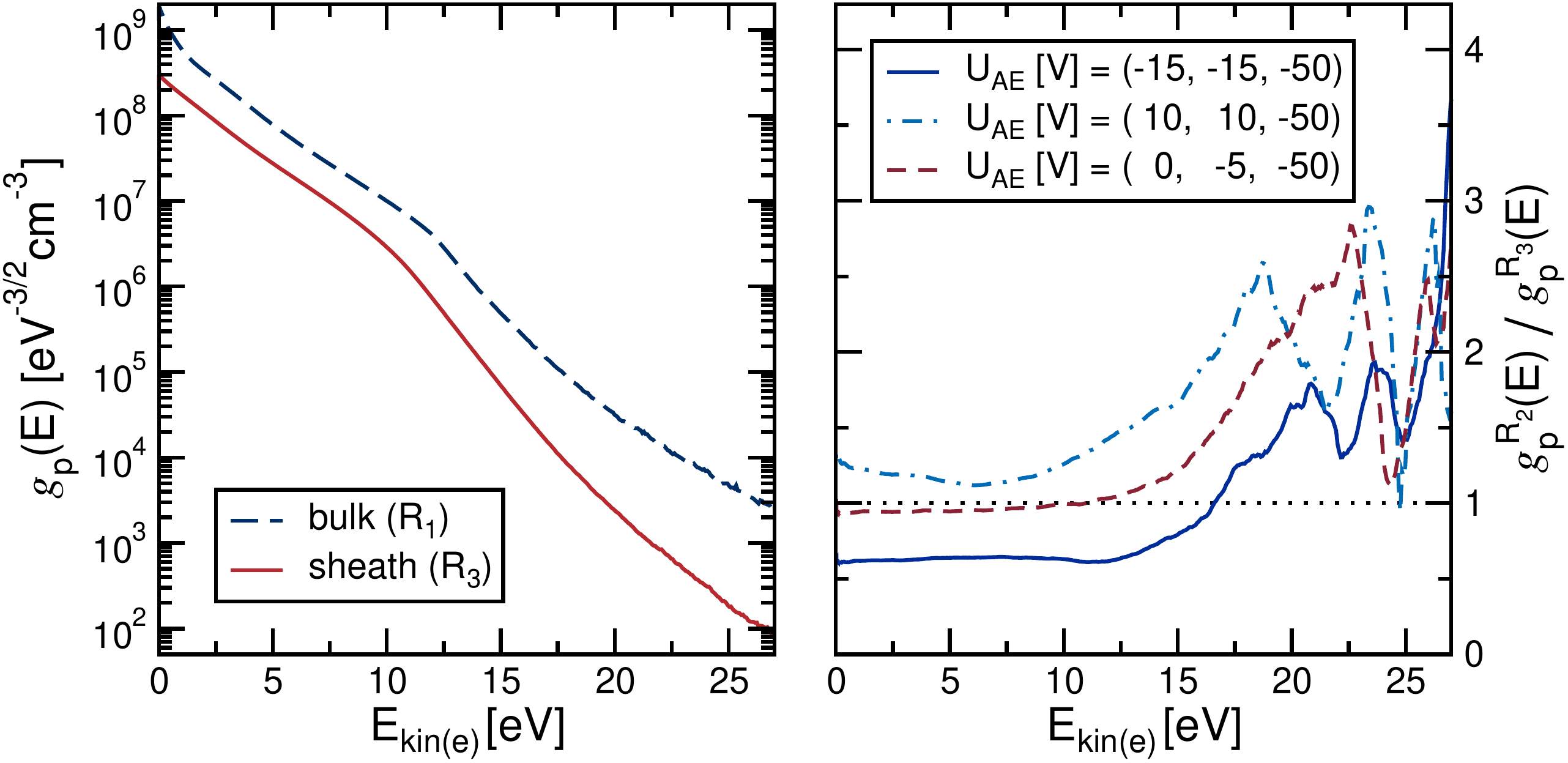}
  \caption{(Color online) Left panel: PIC results for the local
    electron energy probability function ${\textsl{g}}_p(E)$ in the bulk and the
    sheath of the discharge. Right
    panel: Changes in ${\textsl{g}}_p(E)$ above the center AE-pixel induced by
    different biasing.  Away from the biased pixels, the EEPF in
    region $R_3$ agrees for all AE configurations.}
  \label{fig:local_EEPF}
\end{figure}

The density profile does not contain any information on the energy of the
electrons in the sheath.
Therefore, we calculate 
the electron energy probability distribution $g_p(E)$.\cite{LL05}
In Fig.~\ref{fig:local_EEPF} $\textsl{g}_p(E)$ is given for three regions 
of the discharge, representing bulk, unperturbed sheath
and sheath above a biased electrode pixel. 
The scales used in Fig.~\ref{fig:local_EEPF} are chosen such that 
a Maxwellian distribution results in a straight line.
Then the electron temperature is proportional to the inverse slope and
the crossing with the ordinate gives the species density.
The obtained $\textsl{g}_p(E)$ for the bulk closely resembles a bi-Maxwellian
distribution with a larger fraction of cold and and a smaller fraction
of higher energetic electrons.
The latter one is due to those electrons that oscillate between the 
electrodes of the discharge.
In the sheath, the density is reduced for all energies, but most 
pronounced in the high energy tail.
This is due to the retardation caused by the potential difference the
electrons encounter when crossing the sheath.
As compared to the unperturbed sheath, $\textsl{g}_p(E)$ above the center 
pixel clearly reflects the influence of the additional 
AE-biasing (right panel of Fig.~\ref{fig:local_EEPF}).
For energies up to the plasma potential, the density reduction 
(enhancement) for negative (positive) bias is uniform for all energies.
The strong fluctuations at larger energies are due to  
poor statistics since only a tiny fraction of all
electrons have energies in this range.
Irrespective of the biasing on the center pixel, the confinement
potential on the third shell of pixels increases the density in the
high energy tail as compared to the unperturbed sheath through a
channeling effect.

\begin{figure}
\centering\includegraphics[width=\linewidth,clip]{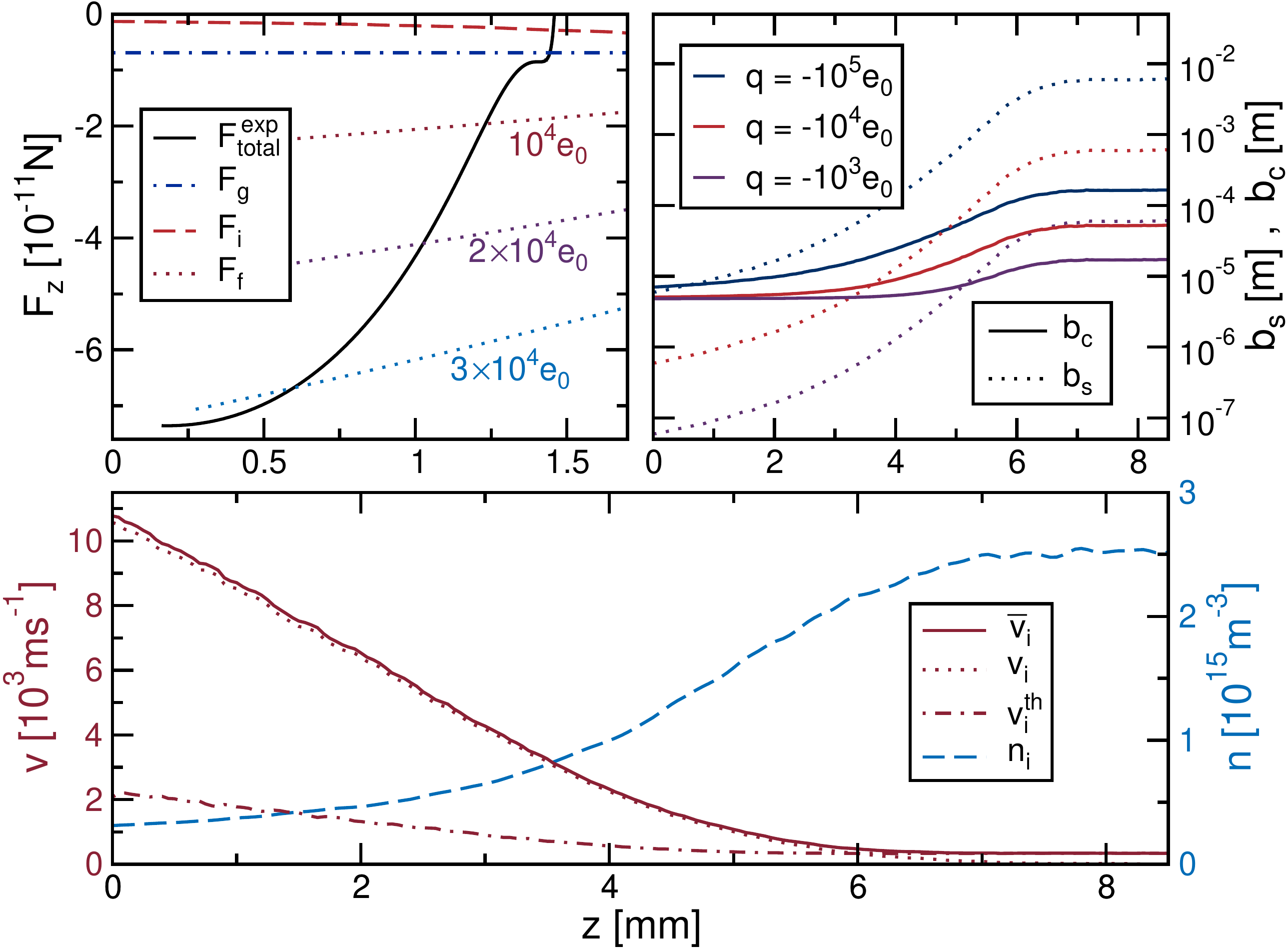}
\caption{(Color online) Upper left panel: Total force
  $F_{\text{total}}^{\text{exp}}$ on the particle as extracted from
  the experimental trajectory. Individual contributions to
  $F_{\text{total}}^{\text{exp}}$ are gravitational force
  $F_{\text{g}}$, ion drag force $F_{\text{i}}$ and electric field
  force $F_{\text{f}}$.  The shown $F_{\text{i}}$ for a particle with
  $q=-10^5e_0$ is an upper bound estimate of this contribution.
  $F_{\text{f}}$ is calculated for various positive particle charges
  using the PIC results for the electric field.  Upper right panel:
  Collection radius $b_{\text{c}}$ and Coulomb scattering radius $b_{\text{s}}$ for
  different particle charges $q$ as derived from the PIC data in the
  lower panel.  Curves and legend correspond from top to bottom.
  Lower panel: PIC results for ion density $n_{\text{i}}$, ion thermal velocity
  $v_{\text{i}}^{\text{th}}$ and ion drift velocity $v_{\text{i}}$ above the center
  pixel of the AE for bias configuration
  $U_{\text{AE}}[\text{V}]=(-20,-20,-50)$.
}
\label{fig:iondrag}
\end{figure}

Explaining the accelerated fall of the particle quantitatively requires
a careful analysis of all relevant forces.
In order to extract the total force on the particle from 
the experimental trajectory, we performed a polynomial fit to the 
measured data (inset of Fig.~\ref{fig:switch}) and calculated its second derivative.
Due to the limited number of data points, the curvature data depend
slightly on the order of the used polynomial, but they agree within 5\% in
the relevant time interval.
Combining trajectory and time dependent force data, the total force on
the particle is given as a function of distance above the AE in the
upper panel of Fig.~\ref{fig:iondrag}.
Gravity is responsible for only about 10\% of the observed particle
acceleration.
Also the ion drag force, which presumably becomes enhanced by the
negative pixel biasing, contributes to the acceleration.
It is given by 
\begin{equation}
  F_{\text{i}} = m_{\text{i}} n_{\text{i}} v_{\text{i}} \bar{v}_{\text{i}}
  (\pi b_{\text{c}}^2 + \pi b_{\text{s}}^2\ln\Lambda)\,,
  \label{eq:ion_drag}
\end{equation}
where $m_{\text{i}}$, $n_{\text{i}}$, $v_{\text{i}}$ are the ion mass,
density and drift velocity.
Accounting for the thermal motion of the ions, we define the effective
velocity $\bar{v}_{\text{i}}^2 = v_{\text{i}}^2 +
(v_{\text{i}}^{\text{th}})^2$, where $v_i^{\text{th}}$ is the thermal
ion velocity.
In~(\ref{eq:ion_drag}), the two contributions to the ion drag force 
correspond to momentum transfer due to collection of ions (index $c$)
and Coulomb scattering (index $s$).
The radii in the associated cross sections are given by~\cite{LL05}
\begin{equation}
   b_{\text{c}} = \frac{d}{2}\sqrt{1-\frac{2 e_0 \Phi_{\text{d}}}
     {m_{\text{i}} \bar{v}_{\text{i}}^2}}
   \quad\text{and}\quad
   b_{\text{s}} = \frac{e_0 q}{2\pi\epsilon_0 m_{\text{i}} \bar{v}_{\text{i}}^2}\,,
   \label{eq:radii}
\end{equation}
where $\Phi_{\text{d}}$ is the potential of the dust particle,
and $\epsilon_0$ the 
permittivity of vacuum.
The divergence of the Coulomb scattering cross section is 
circumvented by using a suitable cutoff beyond which passing
ions are neglected~\cite{LL05,BKFOC92,VC00,KIMT02}.
The shielding effect of the plasma suggests~\cite{VC00} to use the
electron Debye length $\lambda_{\text{D}} =[(\epsilon_0 k_{\text{B}} T_{\text{e}})/(n_{\text{e}} e_0^2)]^{1/2}$
as cutoff, leading to the approximate Coulomb logarithm 
$\ln\Lambda\approx \ln(2\lambda_{\text{D}}/d)$.
Treating the dust particle as a spherical capacitor, its potential
and charge are related by $\Phi_{\text{d}} = q/C$ with 
$C=2\pi\epsilon_0 d\left(1+d/(2\lambda_{\text{D}})\right)$.

Evaluating~(\ref{eq:ion_drag}) requires knowledge about ion density
and velocities in the sheath.
These quantities are directly accessible by PIC and are shown in
the lower panel of Fig.~\ref{fig:iondrag}.
Alternatively, the sheath parameters can be estimated from Child's
law~\cite{LL05}.
%
The large dc-bias applied to the AE pixel locally dominates the
characteristics of the sheath and masks the rf-character of the
global discharge there. 
Low pressure should permit the neglect of ion collisions in the
sheath and due to the additional bias the sheath potential is much
larger than the electron temperature.
Yet, within Child's law the sheath width is less than half the 
width of the PIC solution, with similar density and velocity 
profiles on this reduced scale. 
We attribute this discrepancy to the failure of Child's law to 
account for the local disturbance of the plasma conditions by
the pixel biasing.
Using the PIC data, we may calculate the two contributions to the ion
drag force for different particle charges (see upper right panel of
Fig.~\ref{fig:iondrag}).
The bias switching instantaneously enlarges the sheath and thus the
initial particle position is already deep inside the sheath.
Irrespective of the particle charge, the collection radius~$b_{\text{c}}$
in this region essentially coincides with the particle radius.
Only extremely high charges $(q\sim-10^5e_0)$ can compensate the high
ion velocity resulting in a noticeable second term under the square
root in~(\ref{eq:radii}).
The proportionality of the Coulomb scattering radius~$b_{\text{s}}$ to the
particle charge often results in a dominance of $b_{\text{s}}$ over $b_{\text{c}}$.
In the relevant region, however, the high values of $\bar{v}_{\text{i}}$
result in a small prefactor and $b_{\text{c}}$ and $b_{\text{s}}$ become comparable 
for $q\sim- 5\times10^4e_0$.
Summing up both contributions, the resulting ion drag force for
$q=-10^5e_0$ is given in the upper left panel of Fig.~\ref{fig:iondrag}.
Due to the quadratic dependence of $F_{\text{i}}$ on $b_{\text{c}}$
and $b_{\text{s}}$, the curves for smaller $q$ are marginal on this
scale.
In order to explain the remaining difference to the measured
total force, the only explanation seems to be a positive recharging
of the particle.
The resulting electric field forces for selected positive charges are
given in left panel of Fig.~\ref{fig:iondrag}.
An explanation of the observed behavior requires a
gradual charging to $q\sim 2.5\times10^4e_0$ during the fall.
Crucial for this recharging procedure are:
(i) absence of any electrons in the vicinity of the particle during
the whole rf-cycle, which is guaranteed by the large negative pixel
biasing;
(ii) instantaneous sheath widening and location of the
particle deep inside the widened sheath.
Then the accelerated ions may acquire high enough energies to reach
the particle surface, even for a positively charged particle.

\section{Conclusions}

In this work we studied the behavior of a microparticle in the
plasma sheath upon abrupt changes of the local plasma conditions.
Thereto, we combine experimental studies with theoretical modeling and
numerical simulations.
Local perturbations of the plasma sheath in front of biased segments
of the 'adaptive electrode' are monitored by the behavior of 
microparticles.
Hereby, special emphasis is put on the particle dynamics upon 
bias switching and in the afterglow of the discharge.
Relaxation into a new equilibrium position is a markedly slower
process than the adaption of the particle charge to the local electron
and ion densities.
Whether a particle remains hovering above a biased segment or drops
down depends on the charge equilibrium imposed by the altered plasma
conditions in its vicinity.
Slight changes of the plasma conditions initiate a damped oscillation
of the particle into a new equilibrium position.
Drastically increasing the negative bias of a pixel, locally the
sheath may be deprived completely of electrons.
This out-of-equilibrium situation may disturb the particle's charge
balance to the point of a positive recharging.
The resulting downward force might explain the observed additional
acceleration.
Gravity and ion drag force alone are too weak to account for this
behavior.

\section*{Acknowledgements}

This work has been supported by the Deutsche Forschungsgemeinschaft
under SFB-TR 24, projects A5 and B4. We would like to thank
F.~X.~Bronold, K.~Matyash and R.~Schneider for helpful discussions.

\begin{appendix}

\section{Particle self-force}

\begin{figure}
\centering\includegraphics[height=0.50\linewidth,clip]{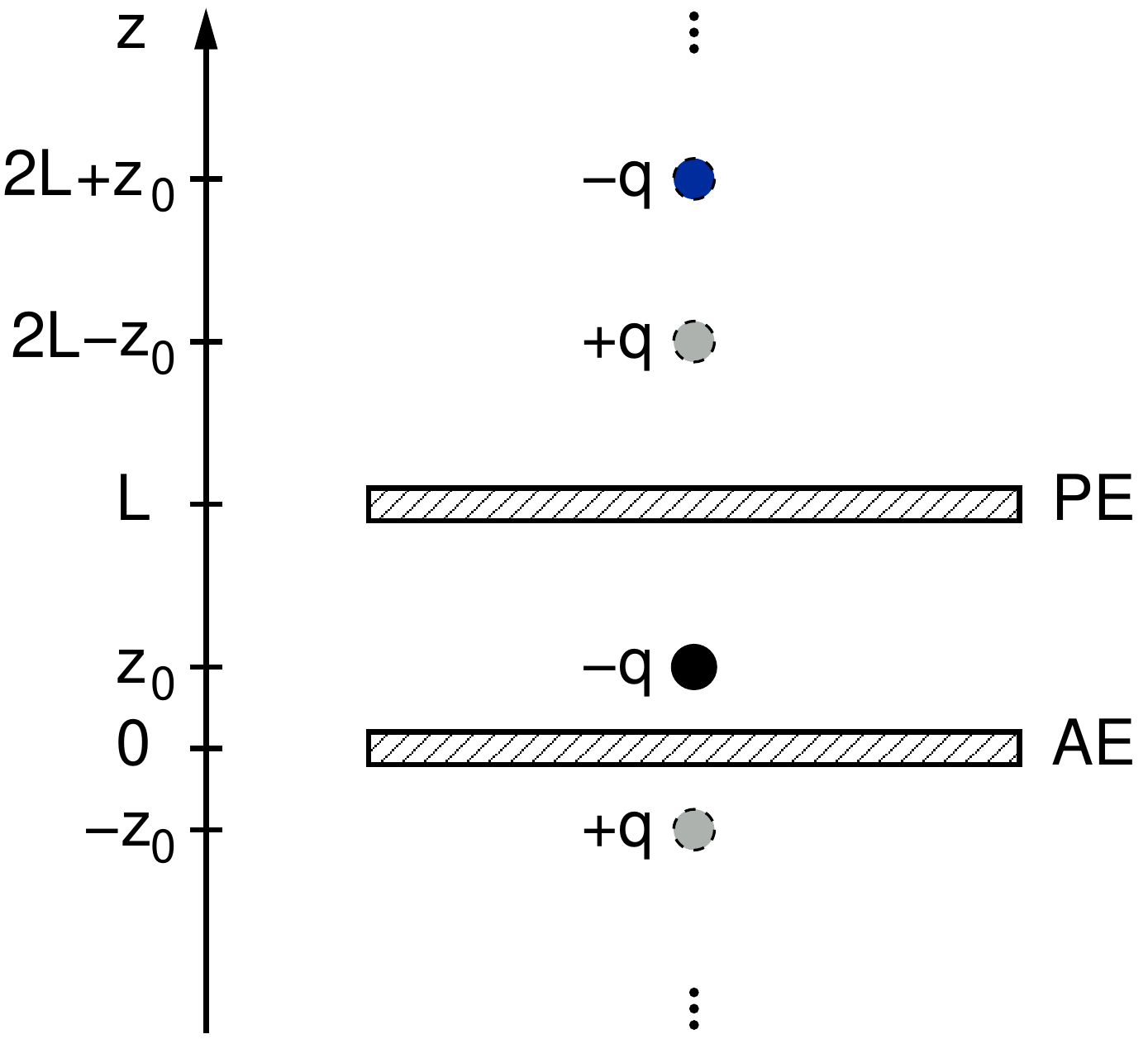}
\hfill
\centering\includegraphics[height=0.50\linewidth,clip]{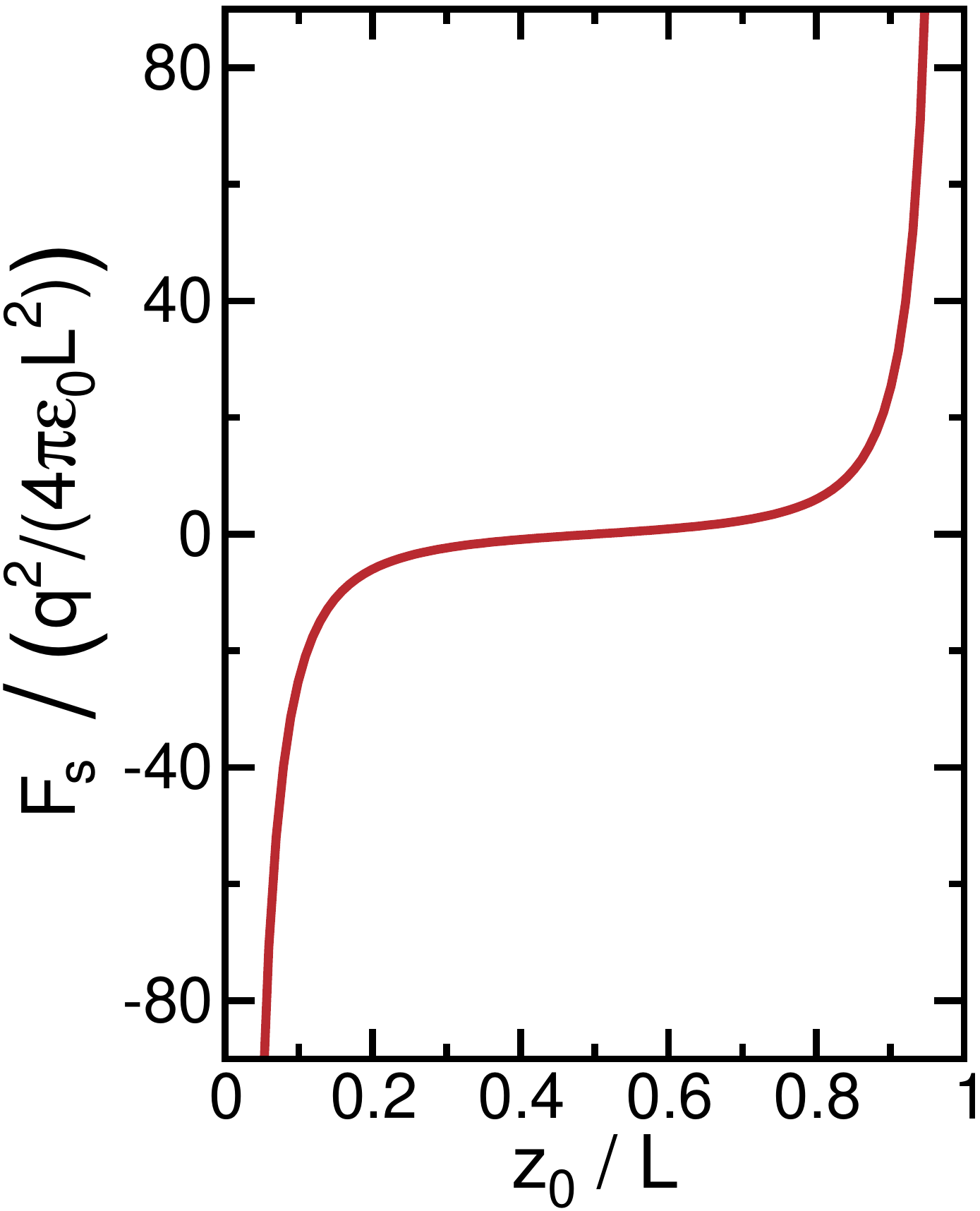}
\caption{(Color online) Left panel: Construction of image charges to calculate
the force on a charged particle (black) between two parallel, grounded electrodes.
Starting with the AE, an image charge of opposite sign (light grey) 
is constructed at $z=-z_0$. To describe the effect of the PE, image charges for
both the original and the first image charge have to be constructed at
$z=2d\pm z_0$. These steps have to be iterated and generate an infinite
series of image charges by repeatedly reflecting the newly created image charges
at the AE and PE. The force acting on the original charge is then
calculated from 
Coulomb's law accumulating contributions from all charges.
Right panel: Force on a particle between two parallel grounded plates as a 
function of distance from the electrode.
}
\label{fig:image_charges}
\end{figure}
Summing up the contributions of all image charges in
Fig.~\ref{fig:image_charges}, we get
\begin{equation}
 F_{\text{s}} = \frac{q^2}{4\pi\epsilon_0}\left(\frac{-1}{4z_0^2} + 
   \sum\limits_{j=0}^{\infty}  \frac{(-1)^{j} }{(z_0-r_j)^2}\right)\,,
 \label{eq:F_plates}
\end{equation}
with $r_j= (-1)^{\lfloor j/2\rfloor} (2L ( \lfloor j/4\rfloor+1) + (-1)^{j+1}z_0)$, where the
floor function $\lfloor.\rfloor$ returns the largest integer that is 
not greater than its argument, e.g., $\lfloor1.75\rfloor=1$.
The infinite sum in (\ref{eq:F_plates}) may be evaluated and with
$x=z_0/L$ and the trigamma function
$\psi_1(x)=\sum_{j=0}^{\infty} 1/(x+j)^2$ we get
\begin{equation}
  F_{\text{s}} =  \frac{q^2}{4\pi\epsilon_0 L^2}
  \left( \frac{\pi^2}{4} \big(1+ \cot(\pi x) \big) -
    \tfrac{1}{2}\psi_1(x )\right)\,.
  \label{eq:F_plates_final}
\end{equation}
In the right-hand panel of Fig.~\ref{fig:image_charges} we show
$F_{\text{s}}$ as a function of distance from the AE.
For the relevant parameters ($L=10\,\text{cm}$, $q\sim 10^4 e_0$) the 
prefactor in (\ref{eq:F_plates_final}) is six orders of magnitude smaller 
than the gravitational force acting on the considered particle.
Thus, $F_{\text{s}}$ is negligible except for the vicinity of
the electrodes at which (\ref{eq:F_plates_final}) diverges.
For the above parameters gravity and $F_{\text{s}}$ become
comparable for $z_0\sim 0.05\,\text{mm}$.

\section{Details on PIC simulation}
\label{App_PIC}

All presented simulation results have been calculated using an electrostatic
$2d(r,z)3v$ PIC code.
Since simulating the whole reactor vessel is infeasible, we restricted 
the simulation volume to the relevant region above the AE
($L_z=5\,\text{cm}$, $r_{\text{max}}=6\,\text{cm}$).
By suitable choice of the substitute discharge parameters we ensured
bulk plasma conditions in accordance with the experimental ones,
$\Phi_{\text{p}} = 21.8\,\text{V}$, $T_{\text{e}}=1.1\,\text{eV}$, $n_{\text{e}}=2.5\times
10^9\,\text{cm}^{-3}$.
While the obtained plasma potential and densities deviate from their
target values less than 1\%, they are slightly larger for the electron
temperature.

Adopting the general code structure and null-collision
method~\cite{Bi91,VS95} from the xpdp2 code suite,~\cite{VBLDR93,VD97}
our code has been tailored to match the specific requirements:
rotational symmetry, inclusion of the AE, individual particle weighting, 
OpenMP-parallelization).
The neglect of Coulomb collisions between charged species can be
justified by the low degree of ionization
($n_{\text{e}}/n_{\text{Ar}}\sim10^{-5}$) for the considered
experimental conditions.
Thermal motion of the fixed background gas is taken fully into account
for ion neutral collisions while being neglected for electron neutral
collisions due to the large mass ratio and thus faint momentum transfer.
The solution of the Poisson equation is performed by means of Fourier accelerated
cyclic reduction~\cite{Sw77} as implemented, e.g., in FISHPACK~\cite{FISHPACK}.
We ensured numerical stability of our results for the considered grid sizes 
($n_z\times n_r = 1000\times 1200$) by using double precision variables for 
the potential and charge density.
The constant grid spacing of $\Delta_r=\Delta_z=0.05\,\text{mm}$
is smaller than half the Debye length.
By choosing a time step of $\Delta t = 2.83\times10^{-11}\,\text{s}$
we ensure $\omega_{\text{p}}\Delta t < 1/5$, where $\omega_{\text{p}}$ is the plasma
frequency.
Respecting these constraints, the PIC method provides reliable
results.~\cite{TMST07}
Modeling a vertical cut through the reactor vessel, symmetry implies a vanishing
electric field at the inner boundary, $r=0$, of the simulation volume. 
At the outer boundary, $r=r_{\text{max}}$, we assume vanishing
electric field boundary conditions since the lateral dimension of the simulation 
volume covers only a fraction of the powered electrode.
In order to model the AE, each pixel has been associated to an individual 
external circuit, and the charge balance is calculated in each iteration.
A linear weighting scheme for particle deposition to the grid and 
interpolation of the electric field ensures conservation of momentum 
and absence of self-forces.
The $r$-dependence of the volume elements in cylindrical coordinates
leads to a drastically varying number of superparticles in each cell when 
describing a constant density.
Alternatively, additional weighting factors for each superparticle may
compensate for this~\cite{SPC08b}.
Thereby, the statistics can be markedly improved since the initial
number of superparticles in all cells is balanced.
Throughout this work we used a mean number of superparticles per cell
of $\langle N_{\text{c}}\rangle\approx80$.
Note that these individual weighting factors have to remain constant
during the whole simulation in order to ensure conservation of energy
and momentum and absence of self-forces.
As a consequence of the experimental setup, the main particle motion is in 
$z$-direction.
Radially, the particles remain within nearby cells of their 
initial position.
Thus, we are in the lucky position that the statistical advantage due
to the individual weighting decays only slowly with time.
In the experiment the AE-biasing does not influence the 
global discharge parameters but only locally disturbs the 
plasma sheath.
This behavior reflects the small ratio of biased electrode to
wall surface.
The restricted size of the simulation volume implies
a more pronounced impact of the AE-pixels on the bulk plasma due to
the larger surface ratio.
Nevertheless, the numerical bulk quantities
with and without AE-biasing differ only by around 1\%.

\end{appendix}


\end{document}